\documentclass[%
 reprint,
showpacs,
amsmath,amssymb,
aps,
]{revtex4-1}

\usepackage{graphicx}
\usepackage{dcolumn}
\usepackage{bm}

\def\ra{\rangle}
\def\la{\langle}
\newcommand{\ii}{\text{i}}

\begin{document}

\preprint{APS/123-QED, RUP-16-18}

\title{Supersymmetry breaking and Nambu-Goldstone fermions in an extended Nicolai model}

\author{Noriaki Sannomiya$^1$\thanks{E-mail address: sannomiya@cams.phys.s.u-tokyo.ac.jp},
Hosho Katsura$^1$, and Yu Nakayama$^{2,3}$}

\affiliation{
$^1$Department of Physics, Graduate School of Science, University of Tokyo, 7-3-1 Hongo, Tokyo 113-0033, Japan\\
$^2$Department of Physics, Rikkyo University, Toshima, Tokyo 177-8501, Japan\\
$^3$Kavli Institute for the Physics and Mathematics of the Universe (WPI),
University of Tokyo, 5-1-5 Kashiwanoha, Kashiwa, Chiba 277-8583, Japan}

\date{\today}
\begin{abstract}
We study a model of interacting spinless fermions in a one dimensional lattice with supersymmetry (SUSY). The Hamiltonian is given by the anti-commutator of two supercharges $Q$ and $Q^\dagger$, each of which is comprised solely of fermion operators and possesses one adjustable parameter $g$. When the parameter $g$ vanishes, the model is identical to  the one studied by Nicolai [H. Nicolai, J. Phys. A: Math. Gen. \textbf{9}, 1497 (1976)], where the zero-energy ground state is exponentially degenerate. On the other hand, in the large-$g$ limit the model reduces to the free-fermion chain with a four-fold degenerate ground state. 
We show that for finite chains SUSY is spontaneously broken when $g > 0$. We also rigorously prove that for sufficiently large $g$ the ground-state energy density is nonvanishing in the infinite-volume limit. 
We further analyze the nature of the low-energy excitations by employing various techniques such as rigorous inequalities, exact numerical diagonalization, and renormalization group method with bosonization. 
The analysis reveals that the low-energy excitations are described by massless Dirac fermions (or Thirring fermions more generally), which can be thought of as Nambu-Goldstone fermions from the  spontaneous SUSY breaking.
\end{abstract}

\pacs{71.10.Fd, 71.10.Pm, 11.30.Pb}
\maketitle


\section{\label{sec:level1}Introduction}
 Recently, Nambu-Goldstone (NG) bosons \cite{PR_Nambu,NC_Goldstone,PR_Goldstone} in non-relativistic systems have drawn renewed attention. This is in part motivated by the clarification of the counting rule for NG bosons proposed in \cite{PRL_Watanabe,PRL_Hidaka}. According to them, non-relativistic NG bosons can be classified into two types, A and B. If we believe in the existence of the effective Lagrangian for the NG	bosons, they live in the coset space of the spontaneously broken symmetries, and type A (B) NG bosons have the second (first) order time derivative such that they are compatible with the coset structure.

In order to relate the classification to the intrinsic symmetry breaking pattern, we consider the vacuum expectation values of commutation relations, $\langle 0|[Q_i,Q_j]|0\rangle$, of generators of broken symmetries, $Q_i$($i=1,2,\cdots, n_{\rm BG}$) in the Hermitian basis. In the effective Lagrangian description, type A NG bosons correspond to the situation in which $\langle 0|[Q_i,Q_j]|0\rangle=0$ for any pairs of $i$ and $j$, while type B NG bosons correspond to the situation in which $\langle 0|[Q_i,Q_j]|0\rangle\neq0$ for some pair of $i$ and $j$. Using the matrix $\rho$ whose elements are defined as\mbox{ $ \rho_{ij}=-\mathrm{i}\langle0|[Q_i,Q_j]|0\rangle$}, the number of NG bosons is given by $n_{\rm BG}-\frac{1}{2}{\rm rank}\rho$.
Here, the number of type A NG bosons is given by $n_{\rm A}=n_{\rm BG}-{\rm rank}\rho$, and the number of type B NG bosons is given by $n_{\rm B}=\frac{1}{2}{\rm rank}\rho$. 
Furthermore, the above mentioned coset structure determines the 
 dispersion relations from the matrix $\rho$. Type A (B) NG bosons have linear (quadratic) dispersion, $\omega\propto|p|$ ($\omega \propto p^2$), where $\omega$ and $p$ are frequency and momentum, respectively. 
The classification is intuitive, but relies on the weakly coupled effective Lagrangian description with the coset structure. In particular, it is not obvious if the argument directly applies to the spontaneous supersymmetry (SUSY) breaking.



The original idea of  SUSY is to combine bosons and fermions in the same representation. The SUSY generators $Q$ and $Q^\dagger$ interchange bosonic and fermionic particles. It was first discovered in the study of the fermionic strings in two-dimension \cite{Gervais:1971ji, Ramond:1971gb}, and then generalized to four-dimensions by Wess and Zumino \cite{Wess_NPB74} (as well as in the earlier work by Golfand and Likhtman \cite{Golfand:1971iw}). In elementary particle physics, SUSY has been pursued for a possible solution of the hierarchy problem \cite{Weinberg:1975gm, Gildener:1976ai}. The central idea is that the Bose-Fermi cancellation above the electro-weak energy scale will ameliorate the high energy fine-tuning of dimensionful parameters in the standard model of particle physics. 

In reality, we have not observed SUSY in any particle physics experiment as of writing this paper. Therefore, we believe that SUSY, if any, must be spontaneously broken. In relativistic field theory, spontaneous SUSY breaking also gives rise to massless fermions, which are called Nambu-Goldstone (NG) fermions or Goldstinos \cite{Witten_NPB82}. This is the SUSY version of the Nambu-Goldstone theorem \cite{PR_Nambu,NC_Goldstone,PR_Goldstone}. Studies on the relationship between NG fermions and spontaneous SUSY breaking have a long history especially in relativistic cases. Indeed, before the seminal paper by Wess and Zumino, Akulov and Volkov \cite{Volkov:1972jx,Volkov:1973ix} proposed the model of SUSY only with fermions, which may be regarded as the action of the NG fermions.


On the other hand, less is known about the relation between spontaneous SUSY breaking and NG fermions in lattice and/or non-relativistic systems.  In condensed matter physics, SUSY is discussed in cold-atom systems, topological insulators, and lattice systems \cite{PRL_Snoek1, PRA_Snoek2, PRA_Lozano, PRL_Yu, PRL_Yu2, PRA_Yu, PRA_Lai, PRA_Blaizot, Science_Grover,PRL_Jian,
JPA_Nicolai, JPA_Nicolai2, PRL03_Fendley,PRL_Fendley_2005, JPA_Fendley1}. Examples of the lattice models with SUSY include Nicolai's model \cite{JPA_Nicolai} and a class of models introduced by Fendley {\it et. al.} \cite{PRL03_Fendley, PRL03_Fendley}. These models are special in that their ground-state degeneracy tends to grow exponentially with system size. 
In the cold-atom context, a possible realization of SUSY in Bose-Fermi mixtures has been theoretically introduced by Yu and Yang \cite{PRL_Yu}. The Hamiltonian describing the system commutes with the fermionic supercharge $Q$ ($Q^\dagger$) that turns a fermion (boson) into a boson (fermion). The system exhibits spontaneous SUSY breaking, i.e., $\la \psi_0| \{ Q, Q^\dagger \} |\psi_0 \rangle \ne 0$ in the ground state $|\psi_0\rangle$ unless the number of particles is zero. 
On the other hand, the dispersion relation of NG fermions in this model is quadratic and can be thought of as type B in the classification scheme by ignoring the difference between commutator and anti-commutator \cite{PRA_Blaizot}. This means that the model satisfies the counting rule for NG bosons despite that the excitation is a fermion. Here, a natural question arises: Can we always apply the counting rule of NG bosons to spontaneous SUSY breaking naively? 

The answer to this question is ``No" because there is a model that may be regarded as an exception of a naive generalization of the counting rules. One of the goals of this paper is to show such an example and to understand the nature of NG fermions induced by the spontaneous SUSY breaking there. More precisely, we introduce the extended version of the Nicolai model, a lattice model with SUSY in one spatial dimension and analyze the vacuum structure and the low energy excitations.  
We show that SUSY is broken spontaneously in this model and the low energy effective field theory is described by a massless Dirac fermion (or Thirring fermion more generally).  In our model, the structure of the vacuum expectation value of broken SUSY generators looks like that of type B NG bosons, i.e. \mbox{$\langle 0 |\{Q,Q^\dagger\}| 0 \rangle\neq0$}, but we find that the dispersion is linear $\omega\propto |p|$ which is supposed to be the dispersion of type A. This suggests that a naive generalization of the counting rules for NG bosons does not apply to the present case. 


The rest of the paper is organized as follows. In Sec. \ref{sec:model}, we define the model and describe the symmetries of its Hamiltonian, including SUSY. The Hamiltonian of the model is defined as the anti-commutator of fermionic superchages $Q$ and $Q^\dagger$.
In Sec. \ref{sec:break}, we first define precisely what we mean by spontaneous SUSY breaking. We then prove that SUSY is spontaneously broken in our model for both a finite chain with $g>0$ and the infinite chain with sufficiently large $g$. In Sec. \ref{sec:NGf}, we study the nature of the low-energy excitations both analytically and numerically. We provide substantial evidence that the low-energy excitations are well described by massless Thirring fermions. 
Concluding remarks are presented in Sec. \ref{sec:conclusion}. In Appendices \ref{sec:free}-\ref{sec:fdp}, we derive some of the formulas used in the main text. In Appendix \ref{sec:Nakayama}, we present an analogue of the NG fermion theorem in translational invariant but not necessarily Lorentz invariant field theories.

\section{Model}
\label{sec:model}
In this section we define the model we study and describe its symmetries. 
We consider a system of spinless fermions on a chain of length $N$. 
Throughout the paper, we assume that $N$ is even and we impose periodic boundary conditions. For each site $j$, we denote by $c^\dagger_j$ and $c_j$ the creation and the annihilation operators, respectively. They obey the usual anticommutation relations
\begin{equation}
\{c_i,c_j^\dagger\}=\delta_{i,j}, \quad \{c_i,c_j\}=\{c_i^\dagger,c_j^\dagger\}=0,
\end{equation}
for all $i,j=1,2, \cdots, N$. As usual, the number operators are defined by $n_j:=c^\dagger_j c_j$. The total fermion number is then defined by $F:= \sum^N_{j=1} n_j$.

\subsection{Hamiltonian and supercharges}
\label{sec:Ham}
The Hamiltonian of our model is defined in terms of the supercharge $Q$ and its Hermitian conjugate $Q^\dagger$ as
\begin{equation}
H = \{ Q, Q^\dagger \}.
\label{eq:Ham}
\end{equation}
In our model, the supercharge $Q$ is made up solely of fermions and is defined by
\begin{equation}
Q := \sum^{N/2}_{k=1} 
( g c_{2k-1} + c_{2k-1}c_{2k}^\dagger c_{2k+1}).
\label{eq:def_Q}
\end{equation}
One can easily verify that $Q$ and $Q^\dagger$ are nilpotent, i.e., $Q^2=(Q^\dagger)^2=0$ by noting that each summand in Eq. (\ref{eq:def_Q}) is nilpotent and anticommutes with the others.  
Without loss of generality, we can assume that $g \ge 0$ because $Q$ with $g \le 0$ can be achieved by local unitary transformations: $c_j \to (-1)^j c_j$. When $g=0$, the supercharge $Q$ becomes identical to the one considered by Nicolai in \cite{JPA_Nicolai}. In this sense, our model is one parameter generalization of the Nicolai model.

\begin{figure}[h]
\includegraphics[width=0.95\columnwidth]{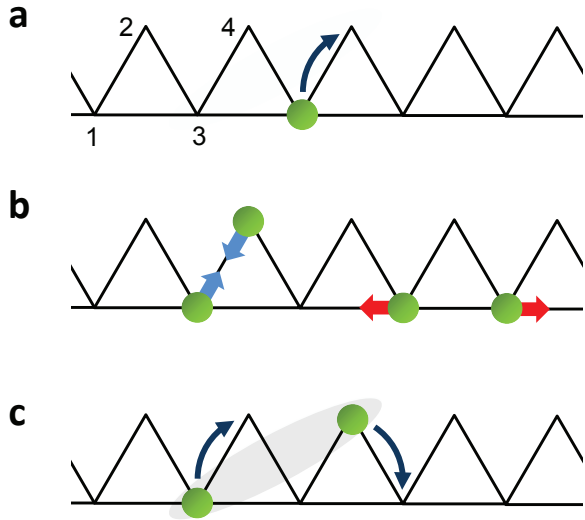}
\caption{(Color online) Schematic representation of (a) the hopping term ($H_{\rm hop}$), (b) the nearest-neighbor attractive and the next-nearest neighbor repulsive interactions ($H_{\rm charge}$), and (c) the pair hopping term ($H_{\rm pair}$). Green (gray) balls represent spinless fermions.}
\label{fig:schematic}
\end{figure}

A tedious but straightforward calculation shows that the explicit expression for the Hamiltonian is given by
\begin{equation}
H = H_{\rm hop}+H_{\rm charge}+H_{\rm pair}+\frac{g^2}{2}N,
\label{eq:Ham2}
\end{equation}
where
\begin{eqnarray}
H_{\rm hop} & = & g\sum_{j=1}^{N}(-1)^j (c_j^\dagger c_{j+1}+c_{j+1}^\dagger c_j),
\label{eq:Hhop}\\
H_{\rm charge} & = & 
\sum^{N/2}_{k=1} (n_{2k} + n_{2k-1} n_{2k+1}) - \sum^N_{j=1} n_j n_{j+1},\label{eq:Hcharge}
\\
H_{\rm pair} & = & \!\sum_{k=1}^{N/2} 
(c^\dagger_{2k}c^\dagger_{2k+3}c_{2k-1}c_{2k+2} +{\rm H.c.}).\label{eq:Hpair}
\end{eqnarray}
The first term $H_{\rm hop}$ denotes the usual fermion hopping term. Here, the alternating sign of the hopping amplitudes is not essential because it can be made uniform by local unitary transformations which will be discussed later. The second term $H_{\rm charge}$ represents an attractive interaction between nearest-neighbor fermions and a repulsive interaction between next-nearest-neighbor fermions on odd-numbered sites. In addition, $H_{\rm charge}$ includes an on-site potential term. The third term $H_{\rm pair}$ looks complicated, but can be thought of as a pair hopping term. A schematic representation of each term is shown in Fig. \ref{fig:schematic}.

\subsection{Symmetries}
\label{sec:sym}
The model has various symmetries including, of course, SUSY. 
It follows from the nilpotency of $Q$ and $Q^\dagger$ that the supercharges commute with the Hamiltonian
\begin{eqnarray}
[H,Q] = [H,Q^\dagger]=0.
\end{eqnarray}
An immediate consequence is that states with positive energy come in pairs, called superpartners. Note that a zero-energy state is a SUSY singlet and does not have a superpartner. 
The supercharges and the total fermion number $F$ satisfy the following algebra
\begin{equation}
[F,Q]=-Q, \quad [ F, Q^\dagger ] = Q^\dagger,
\label{eq:FQ}
\end{equation}
implying $Q$ ($Q^\dagger$) decreases (increases) $F$ by one. As a consequence, $F$ is conserved,
\begin{equation}
[H, F] = 0, 
\end{equation}
from which it follows that the model has U($1$) symmetry. Furthermore, the relation Eq. (\ref{eq:FQ}) tells us that superpartners differ in their fermion number by one. 

Other symmetries of the Hamiltonian are translation, inversion, and charge conjugation. Let us discuss them one by one. The supercharge $Q$ is invariant under translation by two sites, namely, $T: c_j \to c_{j+2}$. As a consequence, $H$ is also invariant under translation $T$. Next, let us examine inversion symmetry. An explicit calculation shows that $Q$ and $Q^\dagger$ are invariant under sending $c_j \to -(-1)^j c_{N-j}$. More precisely, they commute with the operator $U$ that leads to 
\begin{equation}
U c_j U = 
\begin{cases}
-(-1)^j c_{N-j}\quad & j=1,2, ..., N-1 \\
-c_N \quad & j=N.
\label{eq:Uope}
\end{cases}
\end{equation}
Note that $U$ squares to the identity. One can, in principle, derive an explicit expression for $U$ by noting that the operator $P_{i,j} = 1- (c^\dagger_i - c^\dagger_j) (c_i - c_j)$ permutes $c_i$ and $c_j$~\cite{Essler_Hubbard, footnote1}. The fact that $H$ commutes with $U$ follows from 
\begin{equation}
[Q, U ] = [Q^\dagger, U]=0.
\label{eq:QU_com}
\end{equation}
The inversion $U$ plays an important role in the discussion below. 
Finally, we remark that $Q$ and $Q^\dagger$ are interchanged by the charge conjugation: $c_j \to -(-1)^j c^\dagger_j$. As a consequence, the Hamiltonian Eq. (\ref{eq:Ham}) is invariant under this transformation. 


\section{SUSY Breaking}
\label{sec:break}
\subsection{Definition}
\label{sec:def}
Let us first give a precise definition of spontaneous SUSY breaking. The SUSY Hamiltonian Eq. (\ref{eq:Ham}) is, by definition, positive-semidefinite, i.e., 
\begin{equation}
\langle \psi |H |\psi\rangle=\| Q |\psi\rangle \|^2
+\|Q^\dagger |\psi\rangle \|^2 \ \geq 0,
\end{equation}
for any state $|\psi\rangle$. A state annihilated by both $Q$ and $Q^\dagger$ has zero energy, and hence is a ground state of $H$. If a zero-energy ground state exists, SUSY is unbroken. On the other hand, by spontaneous SUSY breaking, we mean that there is no zero-energy state. In other words, the ground-state energy of $H$ is strictly positive. This definition makes perfect sense in any finite volume systems, but is somewhat subtle in the infinite-volume limit, as discussed by Witten~\cite{Witten_NPB82}. This is because SUSY may be restored in the infinite-volume limit if the ground-state energy approaches zero as the system size $N$ increases. To avoid this subtlety, we shall adopt the following definition:\\

\noindent
{\bf Definition:} {\it SUSY is spontaneously broken if the ground-state energy per site is strictly positive}. \\

\noindent
This definition applies to both finite and infinite-volume systems. In the following we first show in our model that SUSY is broken spontaneously in any finite chain when $g>0$. Then we prove that spontaneous SUSY breaking also occurs in the infinite-volume limit when $g$ is sufficiently large. 

\subsection{SUSY breaking in finite chains}
\label{sec:SSB_finite}
Let us prove that for any finite chain SUSY is broken spontaneously when $g > 0$. The key to the proof is to show the existence of an operator $O$ such that $\{ Q, O\}=g$. Let us first show how the spontaneous SUSY breaking follows from the existence of such an operator. Suppose for the sake of contradiction that there exists a zero-energy state $|\psi_0\rangle \neq 0$ annihilated by both $Q$ and $Q^\dagger$. Then, we have
\begin{equation}
\langle \psi_0 | \{Q,O\} |\psi_0\rangle = 0,
\end{equation}
from which it follows that $g=0$. This, however, contradicts the fact that $g$ is nonvanishing. Now the question is if there is an operator $O$ that has the desired property. In fact, a straightforward calculation shows that the following operators
\begin{eqnarray}
O_k = c_{2k-1}^\dagger \bigg[ 1 & - & \frac{1}{g}(c_{2k}^\dagger c_{2k+1}+c_{2k-3}c_{2k-2}^\dagger)\nonumber  \\ 
& + &  \frac{2}{g^2} c_{2k-3}c_{2k-2}^\dagger c_{2k}^\dagger c_{2k+1} \bigg]
\label{eq:cohomo}
\end{eqnarray}
satisfy $\{Q,O_k\}=g$ for all $k=1,\cdots,N/2$. Therefore, there is no zero-energy state and the ground-state energy per site is strictly positive for any finite $N$, implying that SUSY is spontaneously broken. 

We note in passing that SUSY is unbroken in the original Nicolai model corresponding to $g=0$. This is most easily seen by noting that the empty and the fully filled states are annihilated by both $Q$ and $Q^\dagger$ when $g=0$. There are many other zero-energy ground states, the number of which grows exponentially with the system size $N$. This will be discussed elsewhere~\cite{Moriya_inp}.

\subsection{SUSY breaking in the infinite-volume limit}
\label{sec:SSB_infinite}
The above argument does not exclude the possibility that SUSY is restored in the infinite-volume limit, because the ground-state energy per site might become zero. However, for sufficiently large $g$ this is not the case. To prove this, we use Anderson's argument~\cite{PRB_Anderson, PRB_Valent, PRD_Beccaria, PRL_Nie, arXiv_Nie} with which we can obtain a lower bound for the ground-state energy. 

We first note that the sum of the lowest eigenvalues of the terms consisting of $H$ must be equal to or less than the ground-state energy $E_0$. In Eq. (\ref{eq:Ham2}), $H_{\rm charge} + H_{\rm pair}$ is identical to the Hamiltonian of the original Nicolai model ($g=0$) 
and its ground-state energy is zero as discussed in the previous subsection. Thus, we have
\begin{equation}
E_0 \ge E^{\rm hop}_0 + \frac{N}{2}g^2,
\end{equation}
where $E^{\rm hop}_0$ is the ground-state energy of $H_{\rm hop}$. 
Since $H_{\rm hop}$ is a free-fermion Hamiltonian, one can easily compute $E^{\rm hop}_0$ and find $E^{\rm hop}_0 \ge -2gN/\pi$
(see Appendix \ref{sec:free} for the derivation). This yields 
\begin{equation}
E_0 \ge \frac{N}{2} g \left( g - \frac{4}{\pi} \right).
\end{equation}
Therefore, the ground-state energy per site, $E_0/N$, is strictly positive when $g > 4/\pi =1.2732...$. This proves the spontaneous SUSY breaking. 

We remark that the condition $g > 4/\pi$ is sufficient for spontaneous SUSY breaking, but may not be optimal. An improved bound might be obtained by using a more sophisticated decomposition of the Hamiltonian.

\section{Nambu-Goldstone fermions}
\label{sec:NGf}

In this section, we study the nature of the low-energy excitations of our model when SUSY is spontaneously broken. 
In Sec. \ref{sec:var}, based on a variational argument, we prove the existence of low-lying states whose excitation energies are bounded from above by a linear dispersion relation. In Sec. \ref{sec:ED}, we show our numerical results obtained by exact diagonalization. The results provide convincing evidence that the dispersion of the lowest fermionic excitation is linear in momentum $p$ and that the low-energy effective field theory falls into the same universality class as the massless Thirring model. 
To further support this, in Sec. \ref{sec:conti}, we carry out an analysis of the continuum limit of the model using bosonization and renormalization group techniques.

\subsection{Lattice result}
\subsubsection{Variational argument}
\label{sec:var}

In this subsection, we prove that spontaneous SUSY breaking in our model implies the existence of a gapless fermionic excitation. More precisely, we derive a rigorous upper bound for the energy of a low-lying state consisting of states with momenta $\pm p$ relative to the ground state, and show that it is bounded from above by $p$-linear dispersion relation. To this end, we propose a variational ansatz, similar to the Bijl-Feynman ansatz~\cite{PR_Feynman} used in the context of the Heisenberg antiferromagnets~\cite{ZPB_Horsch, PRB_Stringari, JPSJ_Momoi}. In the following, we assume that $g>4/\pi$ so that SUSY is spontaneously broken, and that the ground-state degeneracy is independent of the system size $N$. In fact, our numerical results suggest that the ground-state degeneracy is always four, irrespective of $N$.

Let $|\psi_0\ra$ be a normalized ground state of $H$. 
We can always choose $|\psi_0\ra$ to be annihilated by $Q$~\cite{PRL03_Fendley}. This can be seen as follows. Let $|\psi'\ra$ be an eigenstate of $H$ with energy $E>0$ and suppose $Q|\psi'\ra \ne 0$.  Since $H$ commutes with $Q$ and $Q^\dagger$, 
\begin{equation}
|\psi\ra := |\psi'\ra - \frac{1}{E}Q^\dagger Q|\psi'\ra,
\end{equation}
is also an eigenstate of $H$ with the same energy. It then follows from $Q^2=0$ that $Q|\psi\ra=0$. 
Since $H$ commutes with the total fermion number $F$, the translation $T$, and the inversion $U$, we can further assume that $|\psi_0 \ra$ is a simultaneous eigenstate of $(H,F,T,U)$ with eigenvalues $(E_0, F_0, T_0, U_0)$. Note that possible eigenvalues of $U$ are $\pm 1$, because $U^2=1$. 
Another ground state $Q^\dag|\psi_0 \ra$ is also a simultaneous eigenstate of $(H,F,T,U)$ with eigenvalues $(E_0, F_0+1, T_0, U_0)$, because $[Q^\dag, T]=[Q^\dag, U]=0$. Therefore, the two ground states $|\psi_0 \ra$ and $Q^\dagger |\psi_0\ra$ differ only in their fermion number. 

For our purposes, it is convenient to introduce local supercharges
\begin{equation}
q_k = \frac{g}{2} (c_{2k-1} + c_{2k+1}) + c_{2k-1} c^\dagger_{2k} c_{2k+1},
\end{equation}
which are nilpotent and mutually anticommuting. The Fourier transform of $q_k$ is then defined as 
\begin{equation}
Q_p := \sum^{N/2}_{k=1} e^{-\ii pk} q_k.
\end{equation}
Here, the wavenumber $p$ takes values $p=4 \pi m/N$, where $m \in \mathbb{Z}$ and $-\pi < p \le \pi$. We note that the uniform component $Q_0$ is identical to the supercharge $Q$ in Eq. (\ref{eq:def_Q}). One can verify that for any $p$, 
\begin{equation}
[ F, Q_p ] = -Q_p, \quad [F, Q^\dagger_p] = Q^\dagger_p, 
\end{equation}
hold. This implies that $Q_p$ ($Q^\dagger_p$) decreases (increases) $F$ by one. $Q_p$ and $Q_{-p}$ ($Q^\dagger_p$ and $Q^\dagger_{-p}$) are related to each other by inversion $U$ as follows:
\begin{equation}
UQ_p U = Q_{-p}, \quad UQ^\dagger_p U = Q^\dagger_{-p}.
\label{eq:UQp}
\end{equation}

Now, we consider the following variational state
\begin{equation}
|\psi_p \ra := \frac{(Q_p + Q^\dag_p ) |\psi_0\ra}{\| (Q_p + Q^\dag_p ) |\psi_0 \ra \|}
\quad \quad (p \ne 0).
\end{equation}
(Note that $Q_p + Q^\dag_p$ is Hermitian.) The state $|\psi_p \ra$ is orthogonal to both $|\psi_0\ra$ and $Q^\dagger|\psi_0\ra$ since $|\psi_p \ra$ can be decomposed into a linear combination of states with momenta $\pm p \ne 0$ relative to the ground states. The variational energy of $|\psi_p\ra$ is obtained as 
\begin{align}
\epsilon_{\rm var} (p) & = \la \psi_p | H | \psi_p \ra -E_0 
\nonumber \\
& = \frac{\la \,[ Q_p, [H, Q^\dagger_p]\, ]\, \ra_0}{\la \,\{ Q_p, Q^\dagger_p \}\, \ra_0},
\label{eq:var1}
\end{align}
where $\la \cdots \ra_0$ denotes the ground-state expectation value defined by $\la \cdots \ra_0 := \la \psi_0 | \cdots | \psi_0 \ra$.
A detailed derivation of Eq. (\ref{eq:var1}) is given in Appendix \ref{sec:var1}. 
From the locality of anticommutators,
\begin{equation}
\{ q_k, q^\dagger_\ell \} = 
\begin{cases}
{\rm nonzero} & |k-\ell| \le 1 \\
0 & {\rm otherwise}
\end{cases},
\label{eq:locality}
\end{equation}
and the identity $[H, Q^\dagger_p] = [Q^\dagger, \{ Q, Q^\dagger_p \} ]$, we find that the commutator $[H, Q^\dagger_p]$ is a sum of local operators. However, the double commutator $[ Q_p, [H, Q^\dagger_p]\, ]$ may not be so since $[ q_k, q^\dagger_\ell]$ is nonzero for all $k,\ell$. To obtain a meaningful bound, we use the following inequality~\cite{JLTP_Pitaevskii}
\begin{equation}
| \la \psi |\, [\, A^\dagger, B \,] \,|\psi\ra |^2 \le 
\la \psi|\, \{ A^\dagger, A \} \,|\psi\ra \, \la \psi|\, \{ B^\dagger, B \} \,|\psi \ra
\label{eq:Stringari}
\end{equation}
which holds for any state $|\psi\ra$ and any operators $A$, $B$ (for the proof, see Appendix \ref{sec:Stringari}). With the identification, $|\psi\ra=|\psi_0\ra$, $A=Q^\dagger_p$, and $B=[H, Q^\dagger_p]$,  we have
\begin{equation}
\epsilon_{\rm var} (p)^2 \le 
\frac{\la\, \{\, [Q_p, H], [H, Q^\dagger_p] \,\} \,\ra_0}{\la \,\{ Q_p, Q^\dagger_p \}\, \ra_0}
\label{eq:var2}
\end{equation}

The denominator of Eq. (\ref{eq:var2}) is nonvanishing when SUSY is broken spontaneously. This can be seen as follows. Let us denote by $f_{\rm d}(p):= \la \,\{ Q_p, Q^\dagger_p \}\, \ra_0$ the denominator. $f_{\rm d}(p)$ is of the order of $N$ from the locality Eq. (\ref{eq:locality}). The properties Eq. (\ref{eq:UQp}) and $U^2=1$ imply that $f_{\rm d} (p)$ is an even function of $p$. Then, from $f_{\rm d} (0)=E_0$, we have 
\begin{equation}
f_{\rm d} (p) = N \left( \frac{E_0}{N} + O (p^2) \right), 
\end{equation} 
which ensures that $f_{\rm d} (p)$ is nonvanishing for small enough $p$. 
More precise conditions under which $f_{\rm d} (p)$ is nonvanishing are presented in Appendix \ref{sec:fdp}. Next, let us examine the numerator of Eq. (\ref{eq:var2}). We denote by $f_{\rm n}(p):=\la\, \{\, [Q_p, H], [H, Q^\dagger_p] \,\} \,\ra_0$ the numerator. Using the locality, i.e., Eq. (\ref{eq:locality}), repeatedly, we find $f_{\rm n}(p)$ is a sum of the expectation values of local operators, and hence is of the order of $N$. Again, the properties Eq. (\ref{eq:UQp}) and $U^2=1$ imply that $f_{\rm n} (p)$ is an even function of $p$. This can be seen by noting that
\begin{align}
f_{\rm n} (p) &= \la\, U \{\, [Q_p, H], [H, Q^\dagger_p] \,\} U \,\ra_0 
\nonumber \\
& = \la\, \{\, [Q_{-p}, H], [H, Q^\dagger_{-p}] \,\} \,\ra_0 = f_{\rm n} (-p).
\end{align}
Furthermore, $f_{\rm n}(0)=0$ since $H$ commutes with $Q_0$ and $Q^\dagger_0$. Putting these together, we find
\begin{equation}
f_{\rm n} (p) = N \left( C p^2 + O (p^4) \right),
\end{equation}
where $C$ is a constant independent of $N$. Thus, we arrive at
\begin{equation}
\epsilon_{\rm var} (p) \le \sqrt{\frac{f_{\rm n}(p)}{f_{\rm d}(p)}}
=\sqrt{\frac{C}{E_0/N}}\, |p| + O(|p|^3),
\end{equation}
which gives a rigorous upper bound for the low-lying excitations. 

Some comments are in order. 
(i) The assumption that the ground-state degeneracy is finite and independent of $N$ is crucial in the proof, because otherwise we cannot exclude the pathological cases where all the trial states $|\psi_p\ra$ represent other ground states orthogonal to both $|\psi_0\ra$ and $Q^\dagger|\psi_0\ra$. 
(ii) A field theoretical analogue of the above result can be proved, provided the existence of a local operator analogous to Eq. (\ref{eq:cohomo}). See Appendix \ref{sec:Nakayama} for more details. 
(iii) Since our proof does not rely on the specific form of $q_k$ or the one dimensionality of the lattice, we expect that similar results would hold for more general models in one and higher dimensions. 

\subsubsection{Numerical result}
\label{sec:ED}

\begin{figure}[b]
\includegraphics[width=0.93\columnwidth]{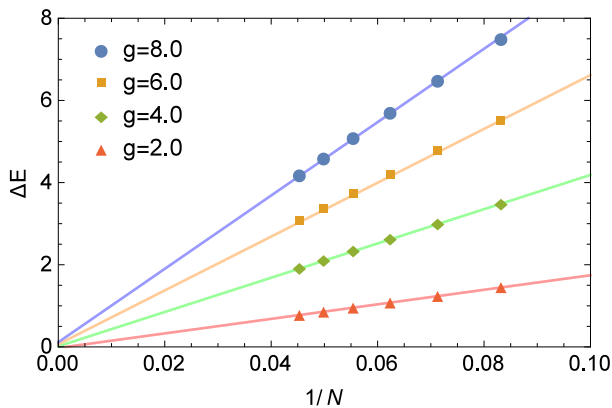}
\caption{\label{fig:epsart} 
(Color online) Finite-size scaling of the lowest excitation energy $\Delta E$ vs $1/N$ for $g=2, 4, 6, 8$. The lines are linear fits to the data. 
}
\label{fig:dispersion}
\end{figure}

The previous argument implies the existence of a gapless excitation with linear dispersion relation. To verify this, we numerically study the low-energy excitations using exact diagonalizaton up to $N=22$ sites. 
We have checked that there are four ground states irrespective of $N$ and they are in the sectors $F=N/2$ and $F=N/2 \pm 1$. The first excited states lie in the subspaces with $F=N/2 \pm 1$ and $F=N/2\pm 2$. 
In Fig. \ref{fig:dispersion}, we plot the first excitation energies relative to the ground state as a function of $1/N$ for various $g$ values. 
Since $1/N$ is proportional to the wavenumber $p$, the dispersion relation can be read off from the plot. The results suggest that the lowest excitation energy is linear in $p$, which implies the existence of NG fermions.

To provide further evidence, we perform a finite-size scaling analysis of the ground state energy density $E_0/N$. From the free-fermion result (Appendix \ref{sec:FSS}) which is valid in the large-$g$ limit, we adopt the following form from the finite size scaling of CFTs \cite{Bloete:1986qm, Affleck_86}: 
\begin{eqnarray}
\frac{E_0}{N}=e_\infty +\frac{\pi v_{\rm F} c}{3N^2}
+O \left( \frac{1}{N^3} \right),
\label{eq:gs_scaling}
\end{eqnarray}
where $e_\infty$ is the ground-state energy per site in the infinite-$N$ limit, $v_{\rm F}$ the Fermi velocity. Here, the constant $c$ is a central charge of the corresponding conformal field theory (CFT). In our numerical analysis, $v_{\rm F}$ is estimated as $v_{\rm F}=N\Delta E/2\pi $, where $\Delta E$ is the difference between the first excited and ground-state energies. 
Combining this estimate with the scaling ansatz Eq. (\ref{eq:gs_scaling}), we find that the central charge for $g=4.0$ is $c=1.00805$, which is remarkably close to unity. The estimated central charges for other $g$ are summarized in Table \ref{tab:c}.
\begin{table}
\caption{The estimated central charge $c$ using Eq. (\ref{eq:gs_scaling})}
\begin{tabular}{c|ccccc}
~~$g$~~ & 2.0 & 4.0 & 6.0 & 8.0\\ \hline
~~$c$~~ & ~0.970468~ & ~1.00805~ & ~1.01969~ & ~1.02534~
\end{tabular}
\label{tab:c}
\end{table}
The results obtained suggest that the low-energy effective field theory of our model with $g>0$ falls into a class of $c=1$ CFTs.

\begin{figure}[h]
\includegraphics[width=0.95\columnwidth]{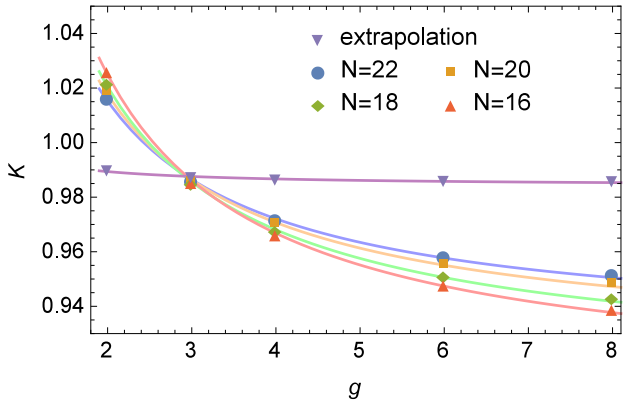}
\caption{(Color online) The TL parameter $K$ as a function of $g$ for $N=16, 18, 20, 22$. The curves are fits to the five data points. As $N$ increases, $K$ converges to a value labeled as extrapolation.}
\label{fig:gvsK}
\end{figure}

\begin{figure}[h]
\includegraphics[width=0.95\columnwidth]{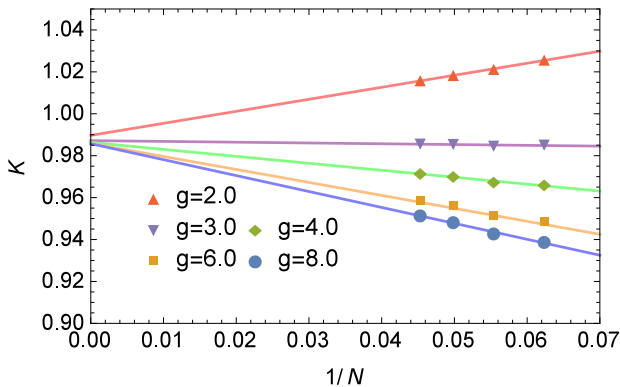}
\caption{(Color online) TL parameter $K$ vs system length $N$ for $g=2, 3, 4, 6, 8$. The lines are fits to the data points. Their intercepts are the TL parameters $K$ in the infinite-size limit.}
\label{fig:NvsK}
\end{figure}

$c=1$ CFTs are further specified by the Tomonaga-Luttinger (TL) parameter (or equivalently, the compactification radius of the dual boson). A well-known example is the massless Thirring model, where the power-law behavior changes continuously with varying the TL parameter. 
To further specify the effective field theory of our model, we calculate the TL parameter numerically based on the formula derived in~\cite{PRB_Hur,JSM_Hur}
\begin{align}
{\mathcal F}(L) & :=\langle[N_A-\langle N_A\rangle]^2\rangle \\
& = \frac{K}{\pi^2}  {\rm log}\left(\frac{{\rm sin}(\frac{\pi L}{N})}{{\rm sinh}(\frac{\pi \alpha}{N})}\right)
\label{eq:nf} 
\end{align}
Here, ${\mathcal F}(L)$ is the number fluctuation in a subsystem $A$ of length $L$, $K$ is the TL parameter, $\alpha$ is a short distance cutoff, $N_A$ is the particle number in subsystem $A$, and the symbol $\langle\cdots\rangle$ denotes the expectation value in the ground state in the sector $F=N/2$. The details of the derivation of the formula are given in Appendix \ref{sec:luttinger}.  
The TL parameter $K$ can be obtained by comparing numerically calculated ${\cal F}(L)$ with Eq.(\ref{eq:nf}).
In this calculation, we take subsystem sizes $L$ from 1 to $N/2$. We note that $K$ takes the same value in the degenerate ground states (in the sector $F=N/2)$. Figure \ref{fig:gvsK} shows the dependence of the estimated $K$ on $g$ for $N=16, 18, 20, 22$. The estimated $K$ increases as $N$ increases when $g$ is larger than about $3.0$, while it decreases when $g$ is smaller than $3.0$. 
Figure \ref{fig:NvsK} shows the estimated $K$ as a function of the system length $N$ for $g=2, 3, 4, 6, 8$. As we can see, $K$ is linear as a function of $1/N$. The intercept of each plot gives the TL parameter of the infinite system. The values of $K$ so obtained are summarized in Table \ref{tab:K} and are shown in Fig. \ref{fig:gvsK} as well. 

\begin{table}
\caption{The TL parameter $K$ in the infinite system estimated by the extrapolation shown in Fig. \ref{fig:NvsK}.}
\begin{tabular}{c|ccccc}
~~$g$~~ & 2.0 & 3.0 & 4.0 & 6.0 & 8.0\\ \hline
~~$K$~~ & ~0.9897 ~ & ~0.9872~ & ~0.9863 ~ & ~0.9858~ & ~0.9857 ~\\ 
\end{tabular}
\label{tab:K}
\end{table} 
According to this table, the TL parameter $K$ is almost independent of $g$ and is remarkably close to unity in the large $N$ extrapolation. This suggests that the low-energy effective field theory is well described by a massless Dirac fermion. In the next section, we compare the numerical results against the renormalization group analysis.
Given a few percent errors in the central charge with nontrivial $g$ dependence (which must vanish) in our extrapolation, however, whether $K=1$ is exactly so independently of $g$ in the infinite-volume limit is to be addressed in a future work.

\subsection{Continuum limit}
\label{sec:conti}

In order to support the numerical results in the previous subsection, we derive the low energy effective Hamiltonian in the large-$g$ limit using bosonization and renormalization group (RG). In this limit, the Hamiltonian $H$ is dominated by $H_{\rm hop}$, and the rest of the terms, $H_{\rm charge}$ and $H_{\rm pair}$, can be treated as a perturbation. 

For later purposes, we rewrite the Hamiltonians in Eqs. (\ref{eq:Hhop})-(\ref{eq:Hpair}) using the following unitary transformation: \mbox{$c_{4j}\!\rightarrow \!-c_{4j}$}, \mbox{$ c_{4j-1}\!\rightarrow\!-c_{4j-1}$}, \mbox{$ c_{4j-2}\!\rightarrow \!c_{4j-2} $}, \mbox{$ c_{4j-3}\!\rightarrow\!c_{4j-3}$}. As a result, we have
\cite{footnote3}
\begin{eqnarray}
H_{\rm hop} & = & -g\sum_{j=1}^{N}(c_j^\dagger c_{j+1}+c_{j+1}^\dagger c_j), \label{eq:hop2}\\
H_{\rm charge} & = & -\sum_{j=1}^N(:\! n_{j}\!:\, :\!n_{j+1}\!:) \nonumber \\ 
& & + \frac{1}{2}\sum_{j=1}^N\left(1+(-1)^j\right)(:\! n_{j-1}\!:\, :\! n_{j+1}\!:), \label{eq:charge2}\\
H_{\rm pair} \! & = &  \!
\sum^{N}_{j=1} \frac{1 \! - \!(-1)^j}{2}
(c^\dagger_{j-2}c_{j-1} c^\dagger_{j+1}c_{j+2} +{\rm h.c.}),
\label{eq:pair2}
\end{eqnarray}
where $:\! n_j \!:$ is normal ordering of $n_j$ which is defined as $n_j-1/2$. We note in passing that one of the ground states is in the sector with $F=N/2$ and the expectation value of $n_j$ in it is $1/2$ due to the translational symmetry.  

Let us first derive a bosonized Hamiltonian. 
In the following, we assume that the parameter $g$ is large enough so that $H_{\rm charge}$ and $H_{\rm pair}$ can be regarded as a perturbation to the free Hamiltonian $H_{\rm hop}$, whose continuum limit is described by free massless Dirac fermions. 
In the continuum limit, the annihilation operator
$\displaystyle c_j$ can be written in the form:
\begin{equation}
c_j \sim \sqrt{a}\left(\psi_L(x)e^{-\ii k_{\rm F}x}+\psi_R(x)e^{\ii k_{\rm F}x}\right).
\end{equation}
Here, $a$ is the lattice spacing, the subscript ${R}$ (${L}$) denotes right (left) branch of fermions, $x=ja$, and $k_{\rm F}=\pi/(2a)$ is the Fermi wave number. Fermion field operators $\displaystyle \psi_L(x)$ and $\displaystyle \psi_R(x)$ satisfy usual anticommutation relations
\begin{eqnarray}
& \{\psi_r(x),\psi_{r'}(y)\} & =\{\psi_r^\dagger(x),\psi_{r'}^\dagger(y)\}=0, \\
& \{\psi_r(x),\psi^\dagger_{r'}(y)\} & =\delta(x-y)\delta_{r,r'},  \quad r,r'=L,R .
\end{eqnarray}
To obtain the effective Hamiltonian in terms of bosonic fields, we bosonize the fermion fields,
\begin{equation}
\psi_R(x)=\frac{1}{\sqrt{2\pi \alpha}}e^{\ii \phi_R(x)} \ , \ \psi_L(x)=\frac{1}{\sqrt{2\pi \alpha}}e^{-\ii \phi_L(x)} ,
\end{equation}
where the bosonic fields $\displaystyle \phi_R(x)$ and $\displaystyle \phi_L(x)$ obey 
the commutation relations
\begin{align}
[\phi_R(x),\phi_R(y)] & =\ii \pi {\rm sgn}(x-y),\\
[\phi_L(x),\phi_L(y)] & =-\ii \pi {\rm sgn}(x-y),\\
[\phi_L(x),\phi_R(y)] & =  \ii \pi,
\end{align}
and the short-distance $\alpha$ is introduced to regularize the ultraviolet divergences. Note that $\alpha$ does not necessarily coincide with the lattice spacing $a$.

\begin{center}
\begin{table}[t]
\caption{Bosonized forms of the interacting terms with their scaling dimensions. For simplicity, we only show the densities of the most relevant terms. 
The scaling dimension of the ${\rm cos}$ term depends on the TL parameter $K$, which is unity in the free-fermion case (large-$g$ limit). The subscript ``uni" indicates the uniform component of each Hamiltonian.}
\label{tab:boson}
\scalebox{0.9}[1.0]{
{\renewcommand\arraystretch{1.5}
\begin{tabular}{ccc}\hline\hline
Hamiltonian & bosonized form & scaling dimension 
\\ \hline \\[-12pt]
$ \sum_j\!:\!n_{j}\!:\, :\!n_{j+1}\!:$
& 
\begin{tabular}{c}
$\displaystyle{\frac{2a^2}{\pi}(\partial_x\varphi)^2}$ \\
$\displaystyle{+\frac{1}{2\pi^2}{\rm cos}(\sqrt{16\pi}\varphi)}$
\end{tabular}
&
\begin{tabular}{c} $2$ \\ $4K$ \end{tabular} 
\vspace{1.5mm}
\\ \hline \\[-12pt]
$(\sum_j\!:\!n_{j+1}\!:\, :\!n_{j-1}\!:)_{\rm uni}$ 
& 
\begin{tabular}{c}
$\displaystyle{-\frac{3a^2}{\pi}(\partial_x\varphi)^2}$ \\
$\displaystyle{-\frac{1}{2\pi^2}{\cos}(\sqrt{16\pi}\varphi)}$
\end{tabular}
& 
\begin{tabular}{c} $2$ \\ $4K$ \end{tabular} 
\vspace{1.5mm}
\\ \hline \\[-12pt]
$(H_{\rm pair})_{\rm uni}$
& 
\begin{tabular}{c}
$\displaystyle{-\frac{15a^2}{8\pi}(\Pi)^2\!+\!\frac{a^2}{8\pi}(\partial_x\varphi)^2}$ \\
$\displaystyle{-\frac{a^2}{2\pi^2\alpha^2}{\cos}(\sqrt{16\pi}\varphi)}$
\end{tabular}
& 
\begin{tabular}{c} $2$ \\ $4K$ \end{tabular} 
\vspace{1.5mm}
\\ \hline\hline
\end{tabular}
}}
\end{table}
\end{center}

The bosonized form of the hopping term reads 
\begin{equation}
H_{\rm hop} \sim 
\frac{v_0}{2} \int dx \{ \, \partial_x\varphi(x)^2 +  \Pi(x)^2 \, \}, 
\label{eq:fb}
\end{equation}
where we have introduced the velocity $v_0=2ga$ and  the new bosonic fields 
\mbox{$
\varphi(x)=(\phi_L(x)+\phi_R(x))/\sqrt{4\pi}
$} and 
\mbox{$
\Pi(x)=(\partial_x\phi_L(x)\!-\!\partial_x\phi_R(x))/\sqrt{4\pi}
$}, which satisfy the commutation relation \mbox{$[\varphi(x),\Pi(y)]=\mathrm{i}\delta(x-y)$}~\cite{Giamarchi}. 
Similarly, $H_{\rm charge}$ and $H_{\rm pair}$ can be expressed in terms of $\varphi$ and $\Pi$. The results are summarized in Table \ref{tab:boson}. We note that higher derivative terms arising from $H_{\rm charge}$ and $H_{\rm pair}$ are negligible since their scaling dimensions are larger than $2$, and thus obviously irrelevant. The staggered terms $\sum_j (-1)^j \cdots$ in Eqs. (\ref{eq:charge2}) and (\ref{eq:pair2}) are also negligible, because in the continuum limit the most relevant terms arising from them can be written in the form of a total derivative. 

Putting all this together, we arrive at the following sine-Gordon Hamiltonian
\begin{eqnarray}
H & \sim & \frac{v}{2}\int dx \left\{  \frac{1}{K}\, \partial_x\varphi(x)^2 + K\, \Pi(x)^2  \right\} \nonumber  \\
& & +\gamma \int dx \ {\rm cos}(\sqrt{16\pi}\varphi(x))
\label{eq:SGham}
\end{eqnarray}
where $v$, $K$, $\gamma$ are the velocity, the TL parameter, and the coupling constant, respectively. 
The values of the velocity and the TL parameter can be read off from Table \ref{tab:boson}: 
\begin{align}
& v=2a\sqrt{\left(g-\frac{15}{8\pi}\right)\!\!\left(g-\frac{27}{8\pi}\right)}, \\
& K\!=\sqrt{\frac{1-15/(8\pi g)}{1-27/(8\pi g)}}. 
\label{eq:ana_K}
\end{align}
In the large-$g$ limit, $K$ is approximated as \mbox{$K \sim 1+3/(4\pi g)$}.
We note that up to this order, the effective action has the emergent Lorentz symmetry, so one may follow the standard procedure of the RG method with Lorentz symmetry. The RG equation for  $\gamma$ is obtained as
\begin{eqnarray}
\frac{d \gamma}{d\ell}\!=\!(2-4K) \gamma,
\end{eqnarray}
where $\ell$ is scale length satisfying $d \ell=d{\rm ln}\alpha$. 
It then follows that the cosine term in Eq. (\ref{eq:SGham}) has scaling dimension $4K>4$ and is obviously irrelevant. Therefore, the low-energy effective Hamiltonian is described by a free-boson theory. The effective Hamiltonian has the same form as Eq. (\ref{eq:fb}), but the velocity $v_0$ is replaced with $v$. By refermionizing the effective Hamiltonian, we can get the Hamiltonian of the massless Thirring model. This supports the conclusion drawn from our numerical results in the previous subsection.

A comment is in order. We saw that in Table \ref{tab:K}, the TL parameter $K$ is almost independent of $g$ and is slightly smaller than $1$ in the large $N$ extrapolation. On the other hand, the analytically obtained $K$ from bosonization in Eq. (\ref{eq:ana_K}) obviously depends of $g$ and is always greater than $1$. This discrepancy may originate from finite-size corrections to Eq. (\ref{eq:nf}) or the regularization scheme chosen in the field theory calculation. While the irrelevance of the $\gamma$-term in Eq. (\ref{eq:SGham}) is not affected by the actual value of $K$ as long as $K$ is sufficiently close to $1$, a more comprehensive analysis of $K$ would be desirable for future studies.  

\section{Conclusion}
\label{sec:conclusion}

In this paper, we have studied one parameter extension of the Nicolai model in one spatial dimension, whose Hamiltonian is constructed from the supercharges $Q$ and $Q^\dagger$ as $H= \{ Q, Q^\dagger \}$. The model interpolates smoothly between the original Nicolai model and the free-fermion chain as the parameter $g$ is varied from $0$ to $\infty$. When $g>0$, SUSY is spontaneously broken for any finite chain, which follows from the existence of a local operator whose anticommutator with $Q$ is constant. For the infinite chain, we proved that SUSY is spontaneously broken when $g>4/\pi$.

We have also carried out various analysis of the nature of the low-energy excitations. Based on a variational approach, we proved the existence of low-lying states whose excitation energies are bounded from above by a linear dispersion relation. We then numerically studied the ground states and the low-lying states to find that the low-energy physics is described by $c=1$ massless Thirring model with the TL parameter $K$ close to $1$.  This was further supported by the analysis of the continuum model in the large-$g$ limit using bosonization and renormalization group techniques.

While we have analyzed a particular model, we can draw some general lessons on NG fermions in non-relativistic systems.
In particular, for the counting of NG fermions, our model provides an example in which assumptions implicitly made in the argument in the literature (for non-relativistic NG bosons) must be carefully reconsidered (see appendix \ref{sec:Nakayama}). They include the assumption on the analytic dependence of the broken SUSY generators on the momentum, decoupling of the NG fermions from the other gapless degrees of freedom and possibilities of strong couplings among NG modes. 
From the RG analysis and the constraint on the spectral functions, we know that relativistic NG modes (both bosonic and fermionic) cannot interact strongly at low energy in two or more space dimensions, but in non-relativistic systems, in particular in $1+1$ dimensions, gapless fermions may admit nontrivial marginal or relevant interactions. In our case, the TL parameter may be an example of such deformations.
To understand the possibility of strong interactions of NG fermions (or bosons more generally without Lorentz invariance), it is important to precisely determine the TL parameter for the NG fermions in our model.

While our RG analysis is essentially $1/g$ expansions around the free Dirac fermions at $g=\infty$, it is interesting to see if we can set up the small $g$ expansions around the original Nicolai model at $g=0$. The Nicolai model has exponentially degenerate ground states, whose classification has not been completed yet \cite{Moriya_inp}, but our argument shows that the perturbation by $g$ lifts all the ground states except for the NG fermion modes. 
In this viewpoint, the Nicolai model may be regarded as a limit of strongly interacting NG fermions and other excitations. We hope that this picture may be helpful to classify the ground states of the original Nicolai model. 

\smallskip

\begin{acknowledgments}
The authors thank Hajime Moriya, Akira Furusaki, and Yutaka Akagi for valuable discussions, and Tohru Koma for bringing our attention to \cite{JLTP_Pitaevskii}. 
HK was supported in part by by JSPS KAKENHI Grant No. JP15K17719 and No. JP16H00985. 
YN was supported in part by the World Premier 
International Research Center Initiative (WPI Initiative), MEXT, Japan.
\end{acknowledgments}

\appendix
\section{Auxiliary free-fermion problem}
\label{sec:free}
\subsection{Ground-state energy}
In this appendix, we calculate the exact ground-state energy of the hopping Hamiltonian Eq. (\ref{eq:Hhop}). The signs of the hopping amplitudes can be made uniform except at the boundary by the following unitary transformation: 
$c_{4m-3} \to c_{4m-3}$, $c_{4m-2} \to c_{4m-2}$, $c_{4m-1} \to -c_{4m-1}$, $c_{4m} \to -c_{4m}$. This yields
\begin{equation}
H_{\rm hop} =
-g \sum^{N-1}_{j=1} (c^\dagger_j c_{j+1}+{\rm H.c.})
-g (e^{\ii \pi N/2} c^\dagger_N c_1 +{\rm H.c.}).
\end{equation}
The boundary conditions are periodic when $N/2$ is even, while they are anti-periodic when $N/2$ is odd. In either case, the Hamiltonian in Fourier space reads
\begin{equation}
H_{\rm hop} = \sum_{q} \epsilon (q)\, {\tilde c}^\dagger_q\, {\tilde c}_q
\quad {\rm with} \quad \epsilon(q) = -2 g \cos q,
\end{equation}
where ${\tilde c}^\dagger_q := \sum_j e^{-\ii qj}c_j/\sqrt{N}$ and the wavenumber $q$ takes the values
\begin{equation}
q = \frac{2\pi}{N} \ell, \quad \quad \ell \in 
\begin{cases}
\mathbb{Z} & \frac{N}{2}:~{\rm even} \\
\mathbb{Z}+\frac{1}{2} & \frac{N}{2}:~{\rm odd}
\end{cases},
\end{equation}
with the constraint $-\pi < q \le \pi$. 

In the ground state of $H_{\rm hop}$, all negative single-particle energy levels are filled. We thus find that the ground-state energy of $H_{\rm hop}$ is
\begin{equation}
E^{\rm hop}_0 = \sum_{-\pi/2 \le q \le \pi/2} \epsilon (q).
\end{equation}
For both periodic and anti-periodic cases, one finds
\begin{equation}
E^{\rm hop}_0 = -\frac{2g}{\tan (\pi/N)}.
\end{equation}
It then follows from the inequality $x \le \tan x$ for $0 \le x \le \pi/2$ that 
\begin{equation}
E^{\rm hop}_0 \ge -\frac{2g}{\pi} N
\label{eq:freegs}
\end{equation}

We note in passing that the ground state of $H_{\rm hop}$ is four-fold degenerate, because $\epsilon (q)$ has two zero modes corresponding to $q=\pm \pi/2$ and the states with these modes filled or empty have the same energy. 

\subsection{Finite-size correction to $E^{\rm hop}_0$}
\label{sec:FSS}

Let us compute the leading finite-size correction to $E^{\rm hop}_0$ (Eq. (\ref{eq:freegs})). For large $N$, the ground-state energy per site behaves as
\begin{align}
\frac{E^{\rm hop}_0}{N} &= 
-2g \left[ \frac{1}{\pi} - \frac{\pi}{3N^2} - \frac{\pi^3 }{45N^4} 
+ O \left(\frac{1}{N^6} \right) \right] \nonumber \\
& = e_{\infty} + \frac{\pi v_{\rm F}}{3N^2} + \cdots,
\end{align}
where $e_{\infty} = -2g/\pi$ is the ground-state energy per site in the infinite-volume limit and $v_{\rm F} = 2g$ is the Fermi velocity. 
The finite size correction of order $1/N^2$ is in agreement with the result of the $c=1$ CFT on a cylinder (e.g. realized by a massless Dirac fermion with periodic boundary condition). 

\section{Derivation of Eq. (\ref{eq:var1})}
\label{sec:var1}

Using a trick similar to the one used in~\cite{ZPB_Horsch}, we have
\begin{align}
\epsilon_{\rm var} (p) & = \frac{\la\, [\, Q_p + Q^\dag_p , [H, Q_p + Q^\dag_p \,] \,] \,\ra_0}{2 \la (Q_p + Q^\dag_p)^2 \ra_0} \nonumber \\
&= \frac{\la\, [\, Q_p + Q^\dag_p, [H, Q_p + Q^\dag_p \,] \,] \,\ra_0}{2 \la \, \{ Q_p, Q^\dagger_p \} \, \ra_0},
\label{eq:var3}
\end{align}
where we have used the fact that $(Q_p + Q^\dag_p)^2 = \{ Q_p, Q^\dagger_p \}$, which follows from the nilpotency of $Q_p$ and $Q^\dagger_p$. 
Since $|\psi_0\ra$ is an eigenstate of the total fermion number $F$ and $Q_p$ ($Q^\dagger_p$) decreases (increases) $F$ by one, the numerator of Eq. (\ref{eq:var3}) can be rewritten as
\begin{align}
   \la \, [\, Q_p, [H, Q^\dagger_p]\, ]\,  \, \ra_0 
+ \la \, [\, Q^\dagger_p, [H, Q_p]\, ]\,  \, \ra_0.
\end{align}
Then, a straightforward calculation shows that
\begin{equation}
\la \, [\, Q_p, [H, Q^\dagger_p]\, ]\,  \, \ra_0 
= \la \, [\, Q^\dagger_p, [H, Q_p]\, ]\,  \, \ra_0,
\end{equation}
yielding Eq. (\ref{eq:var1}).

\section{Proof of the inequality Eq. (\ref{eq:Stringari})}
\label{sec:Stringari}

A proof of Eq. (\ref{eq:Stringari}) can be found in \cite{JLTP_Pitaevskii}. For the readers' convenience, however, we provide a brief proof in this appendix. For notational convenience, we write $\la \cdots \ra := \la \psi | \cdots | \psi \ra$. Using the triangle and the Schwartz inequalities, we have
\begin{align}
& \left| \la \, [ A^\dagger, B ]\, \ra \right|
= \left| \la A^\dagger B \ra - \la BA^\dagger \ra \right|
\nonumber \\
\le & \left| \la A^\dagger B \ra \right| + \left|\la BA^\dagger \ra \right|
\nonumber \\
\le & \sqrt{ \la A^\dagger A \ra \la B^\dagger B \ra }
+ \sqrt{ \la BB^\dagger \ra \la AA^\dagger \ra }.
\end{align}
Then, from the inequality, $2 \sqrt{ab} \le a+b$, we get
\begin{align}
& \left| \la \, [ A^\dagger, B ]\, \ra \right| ^2
\nonumber \\
\le & \la A^\dagger A \ra \la B^\dagger B \ra + \la BB^\dagger \ra \la AA^\dagger \ra 
\nonumber \\
+& 2 \sqrt{\la A^\dagger A \ra \la BB^\dagger \ra \la AA^\dagger \ra \la B^\dagger B \ra}
\nonumber \\
\le & \la \, \{ A^\dagger, A\} \, \ra \la \, \{ B^\dagger, B \} \, \ra,
\end{align}
where we have identified $a$ and $b$ with $\la A^\dagger A \ra \la BB^\dagger \ra$ and $\la AA^\dagger \ra \la B^\dagger B \ra$, respectively. This proves the desired result.\\

\section{A lower bound for $f_{\rm d}(p)$}
\label{sec:fdp}

Similarly to the lower bound for the ground state energy, one finds
\begin{equation}
\{ Q_p, Q^\dagger_p  \} \ge H_{\rm hop} (p) + \frac{N}{4} g^2 (1+\cos p),
\end{equation}
where we write $A \ge B$ to denote that $A-B$ is positive semidefinite. 
The modified hopping Hamiltonian is defined by
\begin{equation}
H_{\rm hop} (p) := \frac{g}{2} \sum^N_{j=1} (-1)^j 
\left[ (1+e^{-\ii p}) c^\dagger_j c_{j+1} +{\rm H.c.} \right]. 
\end{equation}
Again, by a suitable unitary transformation, one can make the hopping amplitudes uniform except at the boundary. The dependence of the ground-state energy on the boundary condition in the free-fermion chain is well known~\cite{PRL_Nie, arXiv_Nie}. It is minimized by the anti-periodic (periodic) boundary condition when $N/2$ is even (odd). In either case, we have
\begin{equation}
H_{\rm hop} (p) \ge -\frac{2g \cos (p/2)}{\sin (\pi/N)}.
\end{equation}
It then follows from the inequality $2x/\pi \le \sin x$ for $0 \le x \le \pi/2$ that $H_{\rm hop} (p) \ge -gN \cos (p/2)$, which yields 
\begin{equation}
\{ Q_p, Q^\dagger_p \} \ge \frac{N}{2} g \cos \frac{p}{2} 
\left( g \cos \frac{p}{2} - 2 \right).
\label{ineq:qqdag}
\end{equation}
The RHS of the above inequality gives a lower bound on $f_{\rm d}(p)$ and is strictly positive for sufficiently small $p$ when $g > 2$, in which case SUSY is broken spontaneously.

\section{Tomonaga-Luttinger parameter and number fluctuation}
\label{sec:luttinger}
In this appendix, we detail the derivation of the formula Eq. (\ref{eq:nf}) obtained in~\cite{PRB_Hur, JSM_Hur}.

\smallskip

A bosonized form of the number density operator is given by
\begin{align}
\rho(x)\sim\rho_0+\frac{1}{\sqrt{\pi}}\partial_x\varphi(x)
\end{align}
where $\rho_0$ is constant and $\varphi(x)$ is the bosonic field introduced in Sec. \ref{sec:conti}. Let $A$ be a segment from $0$ to $l$. Then the number of the fermions in $A$ can be expressed as
\begin{equation}
N_A = \int^l_{0} dx \rho (x).
\end{equation}
It then follows that 
\begin{align}
N_A-\langle N_A \rangle=\frac{1}{\sqrt{\pi}}(\varphi(l)-\varphi(0)), \label{eq:lutt1}
\end{align}
where we have used the fact that $\langle N_A \rangle = \rho_0 l$.
To evaluate the fluctuation of $N_A- \langle N_A \rangle$, it is convenient to introduce the mode expansion of the bosonic field $\varphi(x)$
\begin{align}
\varphi(x)=\varphi_0+\sqrt{\frac{K}{2N}}\sum_{q\neq0}\frac{e^{-\alpha|q|/2}}{\sqrt{|q|}}(e^{\mathrm{i}qx}b_q+e^{-\mathrm{i}qx}b_q^\dagger),
\end{align}
where $b_q$ and $b^\dagger_q$ are, respectively, annihilation and creation operators for a boson with momentum $q$, and $\varphi_0$ is the zero momentum mode. The momentum $q$ takes a value from $\{ \frac{2\pi}{N}n \}_{n \in {\mathbb Z}}$. 
For definitions of $K$ and $\alpha$, see the main text. 
A straightforward calculation yields 
\begin{eqnarray}
\langle[\varphi(l)-\varphi(0)]^2\rangle & = & \frac{K}{2N}\sum_{q\neq0}\frac{e^{-\alpha|q|}}{|q|}(e^{-\mathrm{i}ql}-1)(e^{\mathrm{i}ql}-1) \nonumber\\
& = & \frac{K}{N}\sum_{q>0}\frac{e^{-\alpha q}}{q}(e^{-\mathrm{i}ql}-1)(e^{\mathrm{i}ql}-1) \nonumber\\
& = & \frac{K}{2\pi} \sum_{n=1}^\infty\frac{e^{-\frac{2\pi\alpha}{N}n}}{n}(e^{-\mathrm{i}\frac{2\pi l}{N}n}+e^{\mathrm{i}\frac{2\pi l}{N}n}-2).\nonumber\\
\label{eq:lutt2}
\end{eqnarray}
Then we obtain
\begin{align}
\langle[\varphi(l)-\varphi(0)]^2\rangle & =\frac{K}{2\pi}{\rm log}\left(\frac{{\rm sin}(\frac{\pi}{N}(l-\mathrm{i}\alpha)){\rm sin}(\frac{\pi}{N}(l+\mathrm{i}\alpha)}{{\rm sinh}^2(\frac{\pi\alpha}{N})}\right)\nonumber\\
& \sim\frac{K}{\pi}{\rm log}\left(\frac{{\rm sin}(\frac{\pi l}{N})}{{\rm sinh}(\frac{\pi\alpha}{N})}\right),
\label{eq:lutt3}
\end{align}
where we have assumed that $l$ is much larger than the cutoff length $\alpha$. This together with Eq. (\ref{eq:lutt1}) yields the desired result
\begin{align}
{\mathcal F}(l)=\frac{K}{\pi^2}{\rm log}\left(\frac{{\rm sin}(\frac{\pi l}{N})}{{\rm sinh}(\frac{\pi\alpha}{N})}\right).
\end{align}

\section{NG fermion theorem in field theory}
\label{sec:Nakayama}

\subsection{Existence of gapless fermionic excitation}

In this appendix, we sketch a derivation of the NG fermion theorem in translational invariant  but not necessarily Lorentz invariant field theories.
The argument is almost in parallel with the NG boson theorem \cite{Lange:1966zz, Nielsen:1975hm} modulo some subtleties. In line with the discussions in the main text, we work in the $1+1$ dimensions but higher dimensional generalization is straightforward. (Note that the spontaneous SUSY breaking does occur in $1+1$ dimension, which should be contrasted with the spontaneous bosonic symmetry breaking with Coleman-Mermin-Wagner 
theorem \cite{MW, Coleman}.)

From the locality of the Hamiltonian, we assume that the SUSY generator $Q$ with the SUSY algebra
\begin{align}
Q^2 = Q^{\dagger 2} = 0 , \ \ \{ Q, Q^\dagger \} =  H 
\end{align}
is given by an integral over the local supercharge density $q(x,t)$ as $Q = \int dx q(x,t)$.
We furthermore assume $Q$ is translation invariant $[P,Q] = 0$ with the momentum operator $P$.
Then the conservation of $Q$ requires the existence of another fermionic local operator $q_0(x,t)$ such that $\partial_t q(x,t) = \partial_x q_0(x,t)$. 

In continuum field theory, a well-defined criterion of the spontaneous SUSY breaking is that there exists an local operator $O(x,t)$ and non-zero c-number $g$ such that
\begin{align}
\langle 0 | \{Q, O(x,t) \} | 0 \rangle = g,  
\label{sb}
\end{align}
where $|0\rangle$ is assumed to be translation invariant $P|0\rangle = 0$. 
The operator $O(x,t)$ can be thought of as a continuum analogue of Eq. (\ref{eq:cohomo}) discussed in the main text.

To proceed, we study the anti-commutator $ \{q(x,t) , O(0,0) \}$. For notational convenience, we write $q:=q(0,0)$ and $O:=O(0,0)$ in the following. Using the resolution of the identity in momentum representation, we have
\begin{align}
&\langle 0| \{q(x,t) , O \} | 0 \rangle \cr
&= \sum_{n_p} \int \frac{dp}{2\pi} e^{-\mathrm{i}px + \mathrm{i}E_p t} \langle 0| q| n_p \rangle \langle n_p|O|0 \rangle \cr
&+ e^{\mathrm{i}px - \mathrm{i}E_p t} \langle 0| O| n_p \rangle \langle n_p|q|0 \rangle, \label{qcom}
\end{align}
where $E_p$ denotes the energy difference of the state $|n_p\rangle$ from the vacuum $|0\rangle$. 
We now consider the limit of the integral of \eqref{qcom} over the segment
\begin{align}
&\lim_{\ell \to \infty} \int_{-\ell}^{\ell} dx \langle 0| \{q(x,t) , O \} | 0 \rangle \cr
& = \lim_{p\to 0} \sum_{n_p}  e^{-\mathrm{i}px + \mathrm{i}E_p t} \langle 0| q| n_p \rangle \langle n_p|O|0 \rangle \cr
 &+ e^{ipx - \mathrm{i}E_p t} \langle 0| O | n_p \rangle \langle n_p| q |0 \rangle 
\label{dec}
\end{align}
We assume that the interaction is sufficiently local so that we can neglect the boundary contribution, and the left hand side tends to $\langle 0 | \{Q, O(x,t) \} | 0 \rangle$. Assumption of the spontaneous SUSY breaking \eqref{sb} then demands that the LHS of Eq. (\ref{dec}) is $t$ independent. Therefore, on the right hand side, we must have $\lim_{p\to 0} E_p =0$. Since $g$ is nonzero, both $\langle 0| q| n_p \rangle$ and $ \langle n_p|q|0 \rangle$ cannot simultaneously vanish in the $p\to 0$ limit. Physically, it means that the supercharge density either $q$ or $q^\dagger$ (or both) create a gapless fermionic excitation in the zero-momentum limit.
This is the minimal prediction of the NG fermion theorem without Lorentz symmetry.

Strictly speaking, the above argument does not exclude the possibility that a non-dispersive isolated contribution from a vacuum degeneracy will satisfy \eqref{qcom} without gapless modes. If this were the case, the LHS of \eqref{qcom} would contain a space-time independent constant $c$. However, we may argue that the equal-time anti-commutators of fermionic operators should vanish sufficiently fast in space separation in local field theories, so this possibility is inconsistent with locality. See \cite{Lange:1966zz, Nielsen:1975hm} for more details.

\subsection{How to count NG fermions}
Naively, one may repeat the argument by Nielsen and Chadha \cite{Nielsen:1975hm}  to discuss counting of NG fermions. We, however, point out a couple of subtleties. First of all, in Nielsen-Chadha, the linear (in)dependency of the broken (super)charge plays a significant role. Here, for the {\it SUSY on a finite lattice}, we may always choose $Q^\dag |0\rangle \neq 0$ while $Q | 0 \rangle = 0$ as a definition of the SUSY breaking ground state. (If $Q |0' \rangle \neq 0$ but $Q^\dagger|0' \rangle = 0$, then define $|0\rangle = Q |0'\rangle$.) 
For the moment, suppose this were the case even in the continuum limit. 
Then consider the Fourier transform of \eqref{qcom}
\begin{align}
& \int dx dt\, e^{-\mathrm{i}px + \mathrm{i} \omega t} \langle 0| \{q(x,t) , O \} | 0 \rangle \cr
& = \sum_{n_p} \delta(\omega + E_{-p}) \langle 0| q | n_{-p} \rangle \langle n_{-p}|O|0 \rangle \cr &+ \delta(\omega - E_p)  \langle 0| O | n_p \rangle \langle n_p|q |0 \rangle . \label{Fcom}
\end{align}
Now $Q | 0 \rangle = 0$ means that the second term of the RHS vanishes in the $p \to 0$ limit. Nielsen and Chadha argued that the LHS must be an analytic function of the momentum, and for this to be the case, the dispersion relation must be an even function $E_p = C_2 p^2 + C_4 p^4 + \cdots$. Indeed, this is the mechanism how we would have type B NG bosons. 
However, we have explicitly seen in the main text that the dispersion relation of our NG fermions is linear, so this argument cannot hold. 

If we notice that the low energy excitation of our NG fermion in the continuum limit is a massless Dirac fermion (or weakly coupled Thirring fermion) in the large-$g$ limit, we can immediately realize what went wrong. The assumption $Q^\dag |0\rangle \neq 0$ and $Q | 0 \rangle = 0$  is an ambiguous statement in the field theory limit. Suppose we define $Q^{(\pm)} := \lim_{p \to \pm 0} \int dx e^{\mathrm{i}px} q(x,t)$, then the Dirac sea prescription gives $Q^{(+)\dag} |0\rangle \neq 0$, $Q^{(+)} | 0 \rangle = 0$ but $Q^{(-)\dag}|0\rangle = 0$, $Q^{(-)} | 0 \rangle \neq 0$ since the definition of creation and annihilation flips across $p=0$ (or at the Fermi surface in the lattice model). Once we take this careful limiting procedure, we notice that the Fourier transform \eqref{Fcom} is analytic with the  linear dispersion $E_p =C_1 |p| + \cdots$ because the two non-analytic effects cancel. 
Thus, our NG fermion violates an implicit assumption of Nielsen-Chadha that the (in)dependency of the broken generator is analytic near $p=0$, and eventually it shows the linear dispersion. We emphasize that this is always the case for the relativistic NG fermions.

There is a further issue in counting the number of NG fermions. Even after relaxing the connection between the linear dispersion relations and the linear (in)dependency of the broken generators, NG fermions in our example still seem to violate the counting rule of Nielsen-Chadha (or more recent ones discussed in the introduction) for another reason. The point is that our NG fermion is equivalent to a massless Dirac fermion (or Thirring fermion), and we see that the number of gapless fermionic degrees of freedom is {\it twice} of the number of SUSY that is spontaneously broken. In other words, with the same SUSY algebra and the same SUSY breaking, one can construct a model in which the number of gapless fermionic degrees of freedom is half of our case  (i.e. Majorana). We will discuss such models in a future publication~\cite{Sann}.

Within our model, the reason why we have doubled the number of gapless fermions is because the state that is created by the NG fermion operator $\lim_{p \to 0}  \int dx e^{\mathrm{i}px} q(x,t)$ discussed above shows the kinetic mixing with another gapless fermionic operator. In our lattice model, this is provided by the fermion $c_{2k}$ living in the even site while the NG fermion operator $c_{2k+1}$ sits at the odd site, and they form a Dirac pair in the large-$g$ limit. This violates a hidden assumption in the effective field theory argument in Murayama-Watanabe \cite{PRL_Watanabe} that the NG modes do not couple with other gapless modes. However, in general there is no guarantee this is the case, and we have a concrete counterexample here.



\begin{thebibliography}{99}


\bibitem{PR_Nambu} Y. Nambu and G. Jona-Lasinio,
Phys. Rev. \textbf{122}, 345 (1961).

\bibitem{NC_Goldstone} J. Goldstone,
 Nuovo Cimento \textbf{19}, 154 (1961).

\bibitem{PR_Goldstone} J. Goldstone, A. Salam, and S. Weinberg,
 Phys. Rev. \textbf{127}, 965 (1962).

\bibitem{PRL_Watanabe} H. Watanabe and H. Murayama, 
Phys. Rev. Lett. \textbf{108}, 251602 (2012).

\bibitem{PRL_Hidaka} Y. Hidaka, 
Phys. Rev. Lett. \textbf{110}, 091601 (2013).

\bibitem{Gervais:1971ji} 
  J.~L.~Gervais and B.~Sakita,
  Nucl.\ Phys.\ B {\bf 34}, 632 (1971).

\bibitem{Ramond:1971gb} 
  P.~Ramond,
  Phys.\ Rev.\ D {\bf 3}, 2415 (1971).


\bibitem{Golfand:1971iw} 
  Y.~A.~Golfand and E.~P.~Likhtman,
  JETP Lett.\  {\bf 13}, 323 (1971)
  [Pisma Zh.\ Eksp.\ Teor.\ Fiz.\  {\bf 13}, 452 (1971)].



\bibitem{Weinberg:1975gm} 
  S.~Weinberg,
  Phys.\ Rev.\ D {\bf 13}, 974 (1976).

\bibitem{Gildener:1976ai} 
  E.~Gildener,
  Phys.\ Rev.\ D {\bf 14}, 1667 (1976).

\bibitem{Volkov:1972jx} 
  D.~V.~Volkov and V.~P.~Akulov,
  JETP Lett.\  {\bf 16}, 438 (1972)
  [Pisma Zh.\ Eksp.\ Teor.\ Fiz.\  {\bf 16}, 621 (1972)].
\bibitem{Volkov:1973ix} 
  D.~V.~Volkov and V.~P.~Akulov,
  Phys.\ Lett.\ B {\bf 46}, 109 (1973).

\bibitem{Wess_NPB74} J. Wess and B. Zumino,
 Nucl. Phys. B \textbf{70}, 39 (1974).

\bibitem{Witten_NPB82} E. Witten, 
Nucl. Phys. B \textbf{202}, 253  (1982).

\bibitem{PRL_Snoek1}
M. Snoek, M. Haque, S. Vandoren, and H. T. C. Stoof,
Phys. Rev. Lett. \textbf{95}, 250401 (2005).

\bibitem{PRA_Snoek2}
M. Snoek, S. Vandoren, and H. T. C. Stoof,
Phys. Rev. A \textbf{74}, 033607 (2006).

\bibitem{PRA_Lozano}
G. S. Lozano, O. Piguet, F. A. Schaposnik, and L. Sourrouille,
Phys. Rev. A \textbf{75}, 023608 (2007).

\bibitem{PRL_Yu} Y. Yu and K. Yang,
 Phys. Rev. Lett. \textbf{100}, 090404 (2008).

\bibitem{PRL_Yu2} Y. Yu and K. Yang, 
Phys. Rev. Lett. \textbf{105} 150605 (2010). 

\bibitem{PRA_Yu} T. Shi, Y. Yu, and C. P. Sun,
Phys. Rev. A \textbf{81}, 011604(R) (2010).

\bibitem{PRA_Lai} H-H. Lai and K. Yang
Phys. Rev. A \textbf{91}, 063620 (2015).

\bibitem{PRA_Blaizot} J. P. Blaizot, Y. Hidaka and D. Satow,
 Phys. Rev. A \textbf{92}, 063629 (2015).

\bibitem{Science_Grover} T. Grover, D. N. Sheng and A. Vishwanath,
Science \textbf{344}, 280.

\bibitem{PRL_Jian} S.-K. Jian, Y.-F. Jiang, and H. Yao,
 Phys. Rev. Lett. \textbf{114}, 237001 (2015).

\bibitem{JPA_Nicolai} H. Nicolai,
 J. Phys. A: Math. Gen. \textbf{9}, 1497 (1976).

\bibitem{JPA_Nicolai2} H. Nicolai,
 J. Phys. A: Math. Gen. \textbf{10}, 2143 (1977).

\bibitem{PRL03_Fendley} P. Fendley, K. Schoutens and J. de Boer, Phys. Rev. Lett. \textbf{90}, 120402 (2003).

\bibitem{PRL_Fendley_2005} P. Fendley and K. Schoutens, 
Phys. Rev. Lett. \textbf{95}, 046403 (2005).

\bibitem{JPA_Fendley1} P. Fendley, B. Nienhuis, and K. Schoutens, 
 J. Phys. A: Math. Gen. \textbf{36}, 12399 (2003).

\bibitem{Essler_Hubbard} F. H. L. Essler, H. Frahm, F. G{\"o}hmann, A. Kl{\"u}mper and V. E.  Korepin,
 The one-dimensional Hubbard model  (Cambridge University Press, Cambridge, England, 2005).

\bibitem{footnote1}
More precisely, $U$ can be expressed in terms of two types of operators
$P^{\pm}_{i,j}=1- (c^\dagger_i \pm c^\dagger_j) (c_i \pm c_j)$ that lead to $P^{\pm}_{i,j} c_j P^{\pm}_{i,j} = \mp c_i$. Note that $(P^{\pm}_{ij})^2=1$.

\bibitem{Moriya_inp}
H. Katsura, Y. Nakayama, and H. Moriya, in preparation. 

\bibitem{PRB_Anderson} P. W. Anderson,
 Phys. Rev. \textbf{83}, 1260 (1951).

\bibitem{PRB_Valenti} R. Valent\'i, P. J.Hirschfeld and J. C. Angl\'es d'Auriac,
 Phys. Rev. B \textbf{44}, 3995 (1991).

\bibitem{PRD_Beccaria} M. Beccaria, G. F. De Angelis, M. Campostrini, and A. Feo, Phys. Rev. D \textbf{70}, 035011 (2004).

\bibitem{PRL_Nie} W. Nie, H. Katsura and M. Oshikawa,
 Phys. Rev. Lett. \textbf{111}, 100402 (2013).

\bibitem{arXiv_Nie} W. Nie, H. Katsura and M. Oshikawa, arXiv:1401.2090 [cond-mat.stat-mech] (2014).

\bibitem{PR_Feynman} R. P. Feynman,
 Phys. Rev. \textbf{94}, 262 (1954).

\bibitem{ZPB_Horsch} P. Horsch and W. von der Linden, 
Z. Phys. B \textbf{72}, 181 (1988).

\bibitem{PRB_Stringari} S. Stringari,
 Phys. Rev. B \textbf{49}, 6710 (1994).

\bibitem{JPSJ_Momoi} T. Momoi,
 J. Phys. Soc. Jpn. \textbf{63}, 2507 (1994).

\bibitem{JLTP_Pitaevskii} L. Pitaevskii and S. Stringari,
 J. Low. Temp. Phys. \textbf{85}, 377 (1991).


\bibitem{Bloete:1986qm} 
  H.~W.~J.~Bl\"ote, J.~L.~Cardy and M.~P.~Nightingale,
  Phys.\ Rev.\ Lett.\  {\bf 56}, 742 (1986).

\bibitem{Affleck_86}
I. Affleck, Phys. Rev. Lett. \textbf{56} 746 (1986).

\bibitem{PRB_Hur} H. F. Song, S. Rachel, C. Flindt, I. Klich, N. Laflorencie and K. Le Hur,
 Phys.Rev. B \textbf{85}, 035409 (2012).

\bibitem{JSM_Hur}A. Petrescu, H. F. Song, S. Rachel, Z. Ristivojevic, C. Flindt, N. Laflorencie, I. Klich, N. Regnault and K. L. Hur, J. Stat. Mech. P10005 (2014).

\bibitem{footnote3}
For simplicity, we will assume $N$ is a multiple of four. 

\bibitem{Giamarchi} T. Giamarchi,
 Quantum Physics in One Dimension (Oxford University Press, Oxford, UK, 2004).

\bibitem{Lange:1966zz} 
  R.~V.~Lange,
  Phys.\ Rev.\  {\bf 146}, 301 (1966).

\bibitem{Nielsen:1975hm} 
  H.~B.~Nielsen and S.~Chadha,
  Nucl.\ Phys.\ B {\bf 105}, 445 (1976).



\bibitem{MW}
N. D. Mermin and H. Wagner
Phys. Rev. Lett. \textbf{17}, 1133 (1966).

\bibitem{Coleman}
S. Coleman, Comm. Math. Phys. \textbf{31}, 259 (1973).

\bibitem{Sann}
N. Sannomiya, H. Katsura, and Y. Nakayama, in preparation.

\end{thebibliography}
\end{document}